\documentclass{elsart}
\usepackage{graphics}
\usepackage{float}

\begin{document}
\begin{frontmatter}

\title{Stochastics of Multiple Electron-Photon Head-on Collisions}

\thanks[corr]{Corresponding author.
E-mail: amk@chair12.phtd.tpu.edu.ru}

\author{A. Kolchuzhkin\thanksref{corr}},
\author{A. Potylitsyn, S. Strokov, V. Ababiy}

\address{Tomsk Polytechnic University, Tomsk, Russia, 634034}

\begin{abstract}
The problem of stochastis in multiple electron-photon head-on collisions has 
been considered in this paper. The kinetic equations for the distributions over the 
electron energy and collisions number along with the equations for these
distributions moments have been obtained.
The equations for the first moments have been solved by the iteration method. 
It has been shown that the variance 
of the energy distribution as a function of the mean number of
collisions has a maximum at some value of $\overline n$. 
It is seen from this analysis that 
multiple scattering of electrons influences on the spectra both electrons and 
photons even for the  photon target of small thickness.
The data of approximate analytical calculations agree with the results
of the Monte Carlo simulation.


\end{abstract}


\end{frontmatter}

\section{Introduction}

The problem of laser light  Compton scattering on high-energy electrons
are now considered in the projects relating to the creation of 
$\gamma -\gamma$
colliders, laser-synchrotron sources, laser cooling, diagnostic of 
sub-picosecond
electron bunches and others \cite{Angelo}. It is supposed that intensity of 
laser flash in these problems is so high that an electron can undergo 
several successive collisions passing through a photon bunch 
\cite{Telnov-83,Telnov-95,Telnov-01}. 
The distributions over the electrons energy and collisions number,
the moments of these distributions, and the spectra of scattered 
photons have been studied in this paper using corresponding kinetic equations 
and by statistical simulation methods. It is seen from this analysis that 
multiple scattering of electrons influences on the spectra both electrons and 
photons even for the photon target of small thickness.

\section{Kinetic equations}

Penetration of an electron through a photon target is a stochastic 
process, where both the number of collisions 
of each electron with photons and the energy loss in individual collision are 
random. 

Let us consider an electron with energy $\varepsilon_0$ traveling 
through a bunch of photons with energy $\omega_0$. 
The typical value of the scattering electron angle in the Compton
back-scattering process is determined by the laser photon energy $\omega_0$ 
and doesn't depend on the electron energy $\varepsilon_0$:

\[\overline\theta_e\sim 2\frac{\omega_0}{mc^2}\, , \]

$mc^2$ being the rest energy 
of electron. We shall consider the head-on collisions of laser photons with
energy $\omega_0\sim 1$ eV and electrons with energy $\varepsilon_0\sim 10$ GeV.
In this case the electron deflection angle is much less than characteristic 
radiation angle $mc^2/\varepsilon_0$ and one can neglect the 
electron angular deflection (the straight-ahead approximation). In this
approximation the probability to undergo $n$ collision along the pass 
$l$, $P(n|\varepsilon_0,l)$, obeys the adjoint balance 
equation (the Kolmogorov-Chapman equation) \cite{Feller,We}:

\begin{eqnarray}
P(n|\varepsilon_0,l)=
&(&1-s\Sigma (\varepsilon_0))P(n|\varepsilon_0,l-s)\nonumber \\
&+&s\Sigma (\varepsilon_0) \int _{0}^{\omega_{\mathrm{max}}}
\frac{\Sigma (\omega;\varepsilon_0) }{\Sigma (\varepsilon_0)}
P(n-1|\varepsilon_0-\omega,l)d\omega\, ,
\label{sec:Pn}
\end{eqnarray}

$\Sigma (\varepsilon_0)$ and 
$\Sigma (\omega;\varepsilon_0)$ being the total
and differential macroscopic cross-sections of the Compton scattering, 
$s$ is a small part of $l$, and $\omega$ is the energy of scattered
photon $\displaystyle (0\le \omega\le\omega_{\mathrm{max}}),$

\[\omega_{\mathrm{max}}=\varepsilon_0 \frac{x}{1+x}\]
is the maximum value of the scattered photon energy, 

\[\displaystyle x=\frac{4\omega_0 \varepsilon_0}{(mc^2)^2}\, ,\]

\[\Sigma(\varepsilon_0)=2n_{\mathrm{L}}\sigma(\varepsilon_0)\, ,\]

\[\Sigma (\omega;\varepsilon_0)=
2n_{\mathrm{L}}\frac{d\sigma(\omega;\varepsilon_0)}{d\omega}\, , \]

$\sigma$ and $\displaystyle \frac{d\sigma}{d\omega}$ are the total and 
differential cross-sections,
$n_{\mathrm{L}}$ is the concentration of laser photons in a bunch.

Note that $\Sigma(\varepsilon_0)$ is the 
mean number of collisions of an electron per unit path length and 
$\Sigma (\omega;\varepsilon_0)$ is the mean number of an electron 
collisions with energy loss in unit interval about $\omega$
per unit path length.

The first term in the right side of Eq. (\ref{sec:Pn})
corresponds to electrons which pass the path \( s \) without collisions and
$1-s\Sigma (\varepsilon_0)$ is corresponding probability. These 
electrons have to undergo
\( n \) collisions along the rest path \( l-s. \) The second term 
corresponds 
to the electrons which undergo the first scattering passing the path \( s \) and 
\( s\Sigma (\varepsilon_0) \)
is corresponding probability. These electrons have to undergo \( n-1 \) 
collisions
after that but the energy of electron after first scattering equals  
$\varepsilon_0-\omega$, where
\( \omega \) is random energy of the scattered photon and 
\( \Sigma (\omega;\varepsilon_0)/\Sigma (\varepsilon_0) \)
is the probability density function of $\omega$. In the limit 
\( s\rightarrow 0 \)
Eq. (\ref{sec:Pn}) gives the integro-differential equation for 
\( P(n|\varepsilon_0,l) \) \cite{brems}:

\begin{eqnarray}
\frac{\partial }{\partial l}P(n|\varepsilon_0,l) 
&+& \Sigma (\varepsilon_0)P(n|\varepsilon_0,l)\nonumber \\
&-&\int ^{\omega_{\mathrm{max}}}_{0}\Sigma (\omega;\varepsilon_0)
P(n-1|\varepsilon_0-\omega,l)d\omega=0
\label{dPn}
\end{eqnarray}

with boundary condition

\[P(n|\varepsilon_0,l)|_{l=0}=\delta_{\mathrm{n0}}\, .\]

In a similar way one can obtain the kinetic equation for the probability 
density
function \( P(\varepsilon|\varepsilon_0,l) \) describing the 
energy distribution of electrons after travelling the path $l$:

\begin{eqnarray}
\frac{\partial }{\partial l}&P&(\varepsilon|\varepsilon_0,l)+
\Sigma (\varepsilon_0)P(\varepsilon|\varepsilon_0,l)
-\int ^{\omega_{\mathrm{max}}}_0
\Sigma (\omega;\varepsilon_0)
P(\varepsilon|\varepsilon_0 -\omega,l)d\omega=0
\label{PQ}
\end{eqnarray}

with boundary condition

\[P(\varepsilon|\varepsilon_0,l)|_{l=0}=
\delta (\varepsilon-\varepsilon_0),\]

$\delta(\varepsilon_0-\varepsilon)$ being the Dirac $\delta$-function. 

Eqs. (\ref{dPn}), (\ref{PQ}) can be transformed into the equations for the
moments of distributions 
$P(n|\varepsilon_0,l),~P(\varepsilon|\varepsilon_0,l)$:

\[\overline{n^{\mathrm{k}}}(\varepsilon_0,l)=
\sum\limits_{n=0}^\infty n^{\mathrm{k}} P(n|\varepsilon_0,l),\]

\[\overline{\varepsilon^{\mathrm{k}}}(\varepsilon_0,l)=
\int ^{\varepsilon_0}_{mc^2}
\varepsilon^{\mathrm{k}}P(\varepsilon|\varepsilon_0,l)d\varepsilon.\]

The equation for $\overline n$ and $\overline{\varepsilon^{\mathrm{k}}}$  
has a form

\begin{eqnarray}
\frac{\partial }{\partial l}\overline n(\varepsilon_0,l)+
\Sigma (\varepsilon_0)\overline n(\varepsilon_0,l)
-\int ^{\omega_{\mathrm{max}}}_{0}\Sigma (\omega;\varepsilon_0)
\overline n(\varepsilon_0-\omega,l)d\omega
=\Sigma(\varepsilon_0)\, ,
\label{bar_n}
\end{eqnarray}

\begin{eqnarray}
\frac{\partial }{\partial l}\overline{\varepsilon^{\mathrm{k}}}(\varepsilon_0,l)+
\Sigma (\varepsilon_0)\overline{\varepsilon^{\mathrm{k}}}(\varepsilon_0,l)
-\int ^{\omega_{\mathrm{max}}}_{0}\Sigma (\omega;\varepsilon_0)
\overline{\varepsilon^{\mathrm{k}}}(\varepsilon_0-\omega,l)d\omega=0\, .
\label{Q1}
\end{eqnarray}

The boundary conditions for the moments are

\[\overline n(\varepsilon_0,l)|_{l=0}=0,\]

\[\overline{\varepsilon^{\mathrm{k}}}(\varepsilon_0,l)|_{l=0}=
\varepsilon_0^{\mathrm{k}}.\]

If the relative energy loss of an electron in one collision is small  the 
integro-differential equations (\ref{bar_n}) and (\ref{Q1}) can be
transformed by the Taylor expansion of integrands:

\begin{eqnarray}
\overline n(\varepsilon_0-\omega,l)\approx
\overline n(\varepsilon_0,l)-
\omega\frac{\partial }{\partial\varepsilon_0}
\overline n(\varepsilon_0,l)
+\frac{1}{2}\omega^2\frac{\partial^2 }{\partial\varepsilon_0^2}
\overline n(\varepsilon_0,l)\, ,
\nonumber
\end{eqnarray}

\begin{eqnarray}
\overline{\varepsilon^{\mathrm{k}}}(\varepsilon_0-\omega,l)\approx
\overline{\varepsilon^{\mathrm{k}}}(\varepsilon_0,l)-
\omega\frac{\partial }{\partial\varepsilon_0}
\overline{\varepsilon^{\mathrm{k}}}(\varepsilon_0,l)
+\frac{1}{2}\omega^2\frac{\partial^2 }{\partial\varepsilon_0^2}
\overline{\varepsilon^{\mathrm{k}}}(\varepsilon_0,l).
\nonumber
\end{eqnarray}

This gives the partial differential equations:                                       
\begin{eqnarray}
\frac{\partial }{\partial l}\overline n(\varepsilon_0,l)+
\beta (\varepsilon_0)
\frac{\partial }{\partial \varepsilon_0}
\overline n(\varepsilon_0,l)
-\frac{1}{2}\gamma (\varepsilon_0)
\frac{\partial^2 }{\partial \varepsilon_0^2}
\overline n(\varepsilon_0,l)=\Sigma(\varepsilon_0),
\label{n_csd}
\end{eqnarray}

\begin{eqnarray}
\frac{\partial }{\partial l}\overline{\varepsilon^{\mathrm{k}}}(\varepsilon_0,l)+
\beta (\varepsilon_0)
\frac{\partial }{\partial \varepsilon_0}
\overline{\varepsilon^{\mathrm{k}}}(\varepsilon_0,l)
-\frac{1}{2}\gamma (\varepsilon_0)
\frac{\partial^2 }{\partial \varepsilon_0^2}
\overline{\varepsilon^{\mathrm{k}}}(\varepsilon_0,l)=0.
\label{Q1csd}
\end{eqnarray}

The quantities $\Sigma(\varepsilon_0)$, 
\(\beta (\varepsilon_0) \), and 
\(\gamma (\varepsilon_0) \) in (\ref{n_csd}), (\ref{Q1csd})
are the moments of the macroscopic differential cross-section:

\[
\Sigma (\varepsilon_0)=
\int _{0}^{\omega_{\mathrm{max}}}
\Sigma (\omega;\varepsilon_0)d\omega,
\]

\[
\beta (\varepsilon_0)=
\int _{0}^{\omega_{\mathrm{max}}}\omega\Sigma (\omega;\varepsilon_0)d\omega,
\]

\[
\gamma (\varepsilon_0)=
\int _{0}^{\omega_{\mathrm{max}}}\omega^{2}
\Sigma (\omega;\varepsilon_0)d\omega.
\]

The Eq. (\ref{Q1csd}) for the second moment 
$\overline{\varepsilon^{2}}(\varepsilon_0,l)$ can be 
transformed into the equation for the variance 

\[
\Delta(\varepsilon_0,l)=
\overline{\varepsilon^{2}}(\varepsilon_0,l)-
\overline\varepsilon^2(\varepsilon_0,l)\, .\]

This equation is

\begin{eqnarray}
\frac{\partial }{\partial l}\Delta(\varepsilon_0,l)+
\beta (\varepsilon_0)
\frac{\partial }{\partial \varepsilon_0}
\Delta(\varepsilon_0,l)
-\frac{1}{2}\gamma (\varepsilon_0)
\frac{\partial^2 }{\partial \varepsilon_0^2}
\Delta(\varepsilon_0,l)\nonumber\\
=\gamma (\varepsilon_0)\left(\frac{\partial }{\partial \varepsilon_0}
\overline{\varepsilon}(\varepsilon_0,l)\right)^2.
\label{Delta}
\end{eqnarray}

\section{Cross-sections and related quantities}

The linear Compton scattering differential cross-section is

\begin{eqnarray}
\frac{d\sigma(y;x)}{dy}=\frac{2\sigma_0}{x} \left(1-y+\frac{1}{1-y}
-\frac{4y}{x(1-y)}+\frac{4y^2}{x^2(1-y)^2} \right) ,
\label{compton}
\end{eqnarray}

where $\displaystyle y=\frac{\omega}{\varepsilon_0}, ~\sigma_0=\pi r_0^2$, and
$\displaystyle r_0=\frac{e^2}{(mc^2)^2}$ is the classical radius of electron. 

In the energy region of our interest the invariant dimensionless parameter 
$x$ is small and the integral interaction coefficients
$\Sigma(\varepsilon_0)$, 
\(\beta (\varepsilon_0) \), and 
\(\gamma (\varepsilon_0) \) 

are described by the approximate formulas  

\[
\Sigma (\varepsilon_0)
\approx \Sigma_0\left(1-\frac{\varepsilon_0}{\omega_{\mathrm{p}}}\right) ,
\]

\[
\beta (\varepsilon_0)
\approx \frac{\Sigma_0}{2 \omega_{\mathrm{p}}}\varepsilon_0^2\, ,
\]

\[
\gamma (\varepsilon_0)
\approx \frac{7}{20}\frac{\Sigma_0 }{\omega_{\mathrm{p}}^2}\varepsilon_0^4\, ,
\]

where $\displaystyle \Sigma_0=\frac{16}{3}n_{\mathrm{L}}\sigma_0,~
\omega_{\mathrm{p}}=\frac{(mc^2)^2}{4\omega_0}.$

The quantities $\beta(\varepsilon_0)$ and $\gamma(\varepsilon_0)$ are the mean 
energy loss 
and the mean squared energy loss of an electron per unit path length.

\section{Solution of equations}

The partial differential equations (\ref{n_csd}), (\ref{Q1csd}), and 
(\ref{Delta}) can be solved 
by the iteration method. In the first approximation, where the terms with 
the second derivation are neglected,

\begin{eqnarray}
\overline n(\varepsilon_0,l)=
\Sigma_0 l-2 \log \left(1+\frac{\beta(\varepsilon_0) l}{\varepsilon_0}\right)
\approx \Sigma(\varepsilon_0) l \left(1+
\Sigma_0 l\left(\frac{\varepsilon_0}{2 \omega_{\mathrm{p}}}\right)^2\right)\, ,
\label{n-sol}
\end{eqnarray}

\begin{equation}
\label{eps}
\overline{\varepsilon }(\varepsilon_0,l)=
\frac{\varepsilon_0 }{1+\beta(\varepsilon_0) l/\varepsilon_0}\, ,
\end{equation}

\begin{equation}
\label{var}
\Delta(\varepsilon_0,l)=
\frac{7}{10}\frac{\varepsilon_0^2}{\omega_{\mathrm{p}}}
\frac{\beta(\varepsilon_0) l}{\left ( 1+
\beta(\varepsilon_0) l/\varepsilon_0 \right)^4}\, .
\end{equation}

The second term in Eq. (\ref{n-sol}) is due to increasing of the interaction
cross-section because of electron energy loss in collisions. But it is
seen from the equation that this effect can be neglected in the energy region 
under consideration and the problem can be solved in one-velocity
$\left(\Sigma(\varepsilon_0)=const\right)$ approximation.

It follows from Eq. (\ref{var}) that the variance
\( \Delta(\varepsilon_0,l) \) has a maximum 
at the point, where 

\begin{equation}
\label{Dmax}
\displaystyle \beta(\varepsilon_0) l=\frac{\varepsilon_0}{3}
\end{equation}
and 

\[\displaystyle \frac{\overline \varepsilon(\varepsilon_0,l)}
{\varepsilon_0}=\frac{3}{4}\, .\]

Using Eqs. (\ref{var}) and (\ref{eps}) 
one can derive the formula 

\[\Delta(\varepsilon_0,l)=
\frac{7}{10}\, \frac{\overline\varepsilon ^3}{\omega_{\mathrm{p}}}
\left(1-\frac{\overline \varepsilon}{\varepsilon_0}\right),\]

which was earlier obtained in \cite{Telnov-97}. 

It can be shown that in the one-velocity approximation the solution of 
Eq. (\ref{PQ}) is the sum of the terms 
corresponding to nonscattered and scattered electrons:

\[P(\varepsilon|\varepsilon_0,l)=
\exp\left(-\Sigma(\varepsilon_0)l\right) \delta(\varepsilon_0-\varepsilon)+
\tilde P(\varepsilon|\varepsilon_0,l)\, , \]

\[\tilde P(\varepsilon|\varepsilon_0,l)=
\sum\limits_{n=1}^\infty 
P_{\mathrm{n}}(\varepsilon_0,l) U_{\mathrm{n}}(\varepsilon|\varepsilon_0)\, ,\]

where

\[P_{\mathrm{n}}(\varepsilon_0,l)=\exp\left(-\Sigma(\varepsilon_0)l\right)
\frac{\left(\Sigma(\varepsilon_0\right)l)^{\mathrm{n}}}{n!}\]

is the Poisson distribution with the mean value 
$\overline n(\varepsilon_0,l)=\Sigma(\varepsilon_0) l$ and the function
$U_{\mathrm{n}}(\varepsilon|\varepsilon_0)$ obeys the recurrent convolution 
formula

\[U_{\mathrm{n}}(\varepsilon|\varepsilon_0)=
\int\limits_0^{\omega_{\mathrm{max}}}
w(\omega;\varepsilon_0)
U_{\mathrm{n-1}}(\varepsilon|\varepsilon_0-\omega)d\omega\, ,\]

where

\[w(\omega;\varepsilon_0)=
\frac{\Sigma(\omega;\varepsilon_0)}{\Sigma(\varepsilon_0)}\]

is the probability density function of $\omega$ and
\[U_1(\varepsilon|\varepsilon_0)=w(\varepsilon_0-\varepsilon;\varepsilon_0)\, .\]

It should be pointed out that $U_{\mathrm{n}}(\varepsilon|\varepsilon_0)$ is the 
energy distribution of n-scattered electrons.

Decreasing of electrons energy due to their repeated collisions changes the 
energy spectrum of scattered photons. The photons produced in the secondary
Compton scatterings are softer and the resulting spectrum can be written as 
the sum of terms corresponding to individual collisions:

$$\Phi(\omega|\varepsilon_0,l)=
\sum\limits_{k=1}^\infty P_k(\varepsilon_0,l)
\sum\limits_{n=1}^k \Phi(\omega|n),$$

where $\Phi(\omega|n)$ is the spectrum of photons resulting from the $n-$th
electron collision and can be written in the form

\[
\Phi(\omega|n)=
\int\limits_{\omega}^{\varepsilon_0}~w(\omega;\varepsilon)~
U_{\mathrm{n-1}}(\varepsilon|\varepsilon_0)d\varepsilon \, .
\]


\section{Numerical results}

The results of analytical calculations above agree with the data of our 
statistical simulation for 10 GeV electrons and 1 eV photon head-on collisions. 
It was supposed in this simulation that the number
of electron collisions with laser photons is random. This number was 
selected from
the Poisson distribution with fixed \( \overline n \). The simulation of 
individual collisions was carried out in the electron rest frame 
using the Klein and Nishina formula with the Lorentz
transformation to the lab system. In the same way as in analytical 
calculations above we neglected the angular deflection of electrons but 
accounted for the energy decreasing after each collision. All results were 
obtained with statistics more than $10^6$ trajectories.

\begin{figure}[H]
\centering
\includegraphics{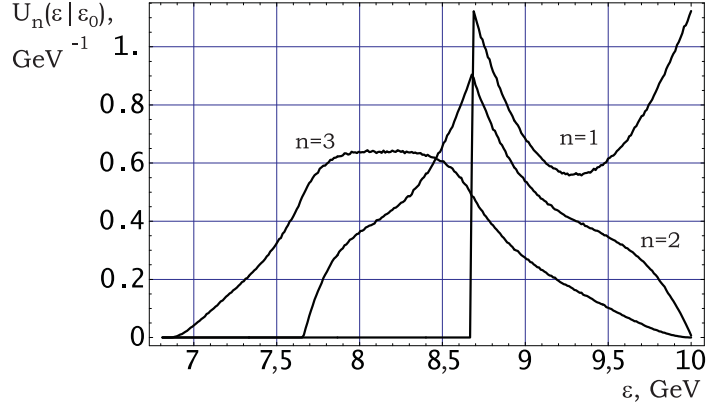}
\caption{Energy spectra of electrons after $n$ = 1, 2, 3 collisions. 
$\varepsilon_0=10GeV, \, \omega_0=1eV.$}
\label{fig:Un}
\end{figure}

\begin{figure}[H]
\centering
\includegraphics{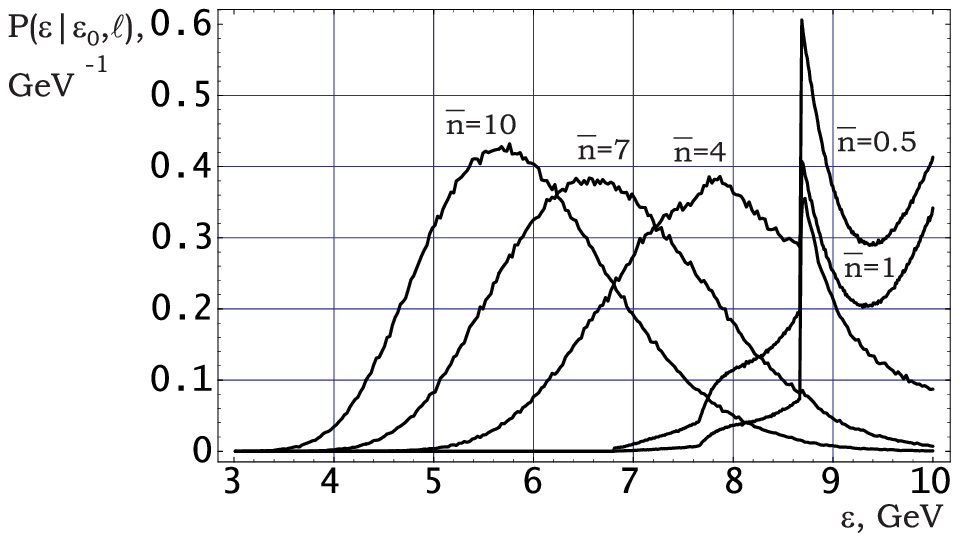}
\caption{Energy spectra of electrons for $\overline n$ = 0.5, 1, 4, 7, 10. 
$\varepsilon_0=10GeV, \, \omega_0=1eV.$}
\label{fig:pq}
\end{figure}

The energy spectra of electrons $U_{\mathrm{n}}(\varepsilon|\varepsilon_0)$ 
after fixed number of collisions are given in Fig. (\ref{fig:Un}).
The data on the electrons energy distributions 
$P(\varepsilon|\varepsilon_0,l)$
are given in Fig. (\ref{fig:pq}) for several values of $\overline n.$ 

It should be pointed out the discontinuous of the spectra for small 
$\overline n$ at the point 
$\varepsilon=\varepsilon_0-\omega_{\mathrm{max}}$ due to single scattered electrons and
contribution of multiple scattered electrons at energies below the point of
discontinuous. This contribution exists even for small $\overline n$.

It is seen from Fig. (\ref{fig:pq}) that the width of the electron energy 
distributions
decreases for such $\overline n$ which are greater than those one determined by 
Eq. (\ref{Dmax}).

\begin{figure}[H]
\centering
\includegraphics{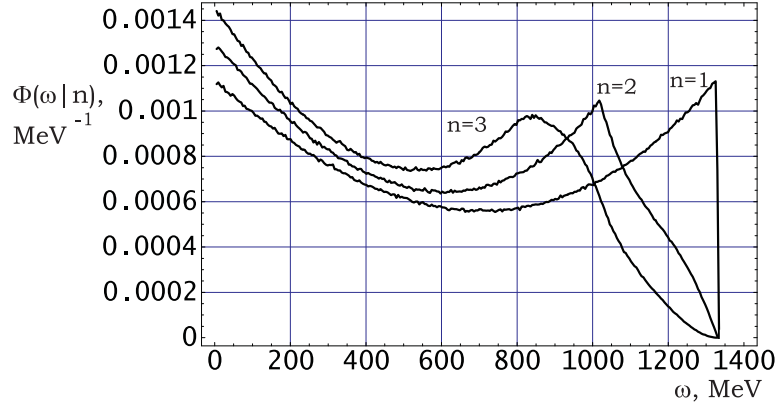}
\caption{Energy spectra of photons generated in the $n$-th collision, $n$ = 1, 2, 3. 
$\varepsilon_0=10GeV, \, \omega_0=1eV.$}
\label{fig:Phi-n}
\end{figure}

\begin{figure}[H]
\centering
\includegraphics{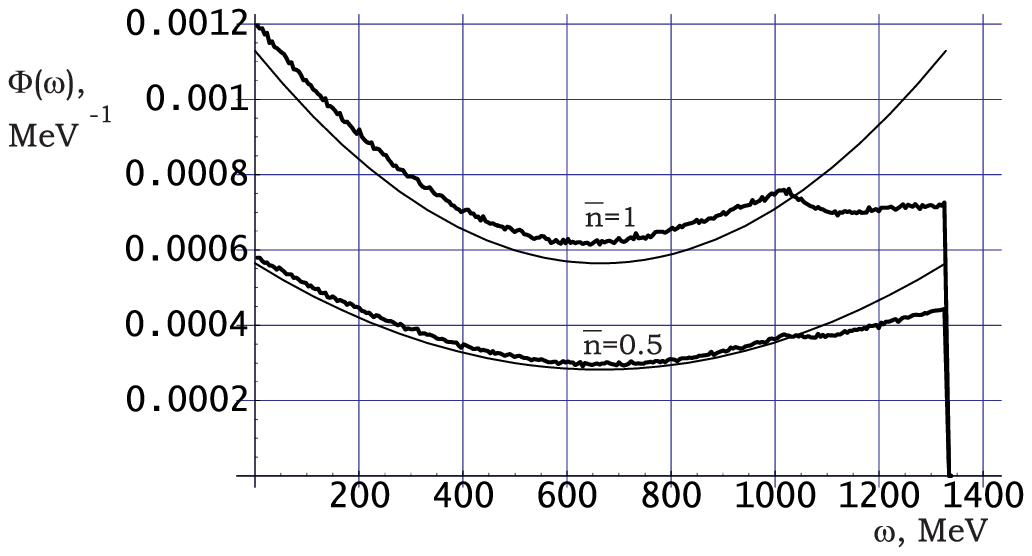}
\caption{Energy spectra of photons for $\overline n$ = 0.5, 1. 
$\varepsilon_0=10GeV, \, \omega_0=1eV.$}
\label{fig:Phi}
\end{figure}

The energy spectra of photons $\Phi(\omega|n)$ from the $n-$th 
scattering of electron and the resulting spectra for several $\overline n$ are 
shown in Fig. (\ref{fig:Phi-n}) and Fig. (\ref{fig:Phi}). It is seen from 
Fig. (\ref{fig:Phi}) that the contribution of multiple scattering should be 
taken into account even for the photon target of small thickness.

\begin{figure}[H]
\centering
\includegraphics{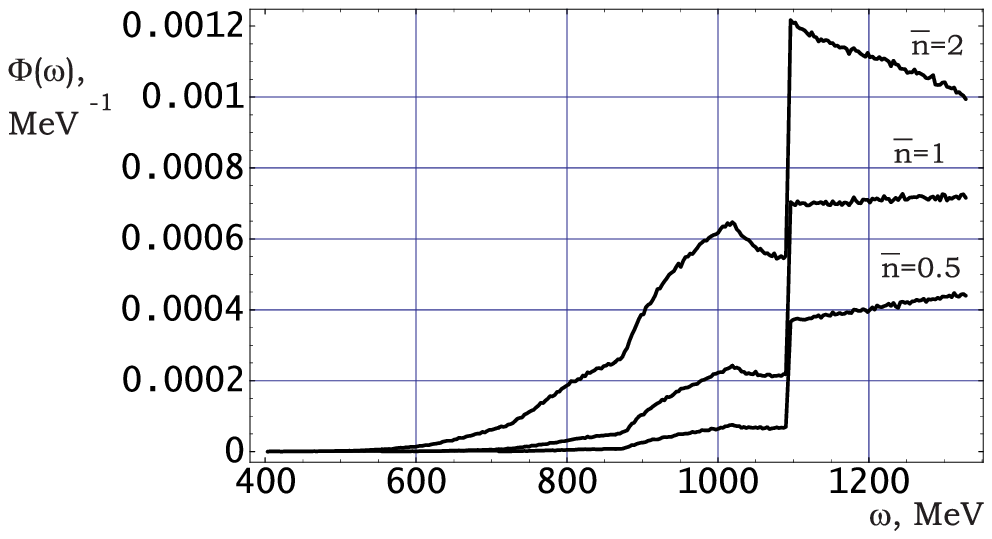}
\caption{Energy spectra of photons with emission angle 
$\displaystyle \theta_{\mathrm{\gamma}} < \frac{1}{2} \, \frac{mc^2}{\varepsilon_0}$
for $\overline n$ = 0.5, 1, 2. 
$\varepsilon_0=10GeV, \, \omega_0=1eV.$}
\label{fig:collimator}
\end{figure}

Fig. (\ref{fig:collimator}) shows that the electrons multiple collisions
influence on the spectra of photons 
with emission angle 
$\displaystyle \theta_{\mathrm{\gamma}} < \frac{1}{2} \, \frac{mc^2}{\varepsilon_0}$.
It is seen from the figure that multiple Compton
scattering results in broadening of the photon spectra with increasing of the 
photon target thickness.

\begin{figure}[H]
\centering
\includegraphics{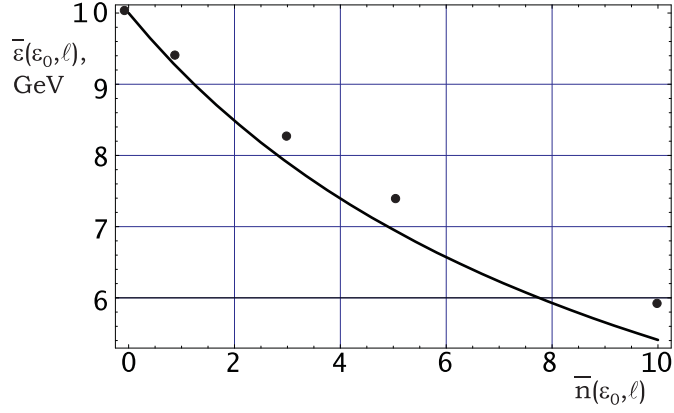}
\caption{Mean energy of electron in back Compton scattering. 
$\varepsilon_0=10GeV, \, \omega_0=1eV.$
Points - simulation, solid line - Eq. (\ref{eps}).}
\label{fig:mean-E}
\end{figure}

\begin{figure}[H]
\centering
\includegraphics{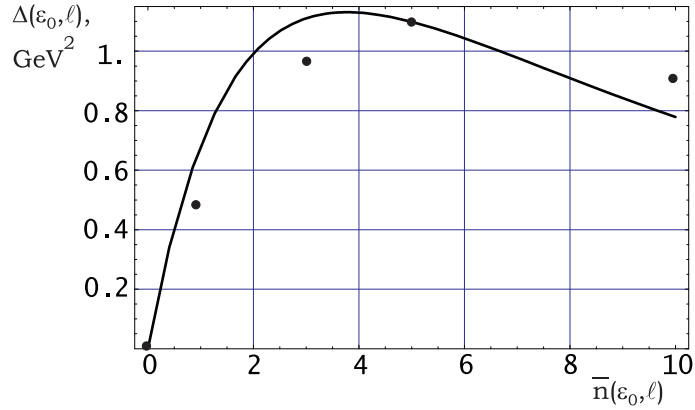}
\caption{Variance of electron energy distribution.
$\varepsilon_0=10GeV, \, \omega_0=1eV.$
Points - simulation, solid line - Eq. (\ref{var}).}
\label{fig:Variance}
\end{figure}

The mean energy loss of electrons in backward Compton scattering and the variance
of the electron energy distribution are shown in Fig. (\ref{fig:mean-E}) and
Fig. (\ref{fig:Variance}). 

It is seen that the Monte Carlo data agree with
the analytical calculations above.

\end{document}